\documentclass[12pt]{article}

\begin{document}

\input amssym.tex

\title{Dirac-K\" ahler equations on curved spacetimes}

\author{Ion I. Cot\u aescu \thanks{E-mail:~~cota@physics.uvt.ro}\\
{\small \it West University of Timi\c soara,}\\
{\small \it V. Parvan Ave. 4 RO-300223 Timi\c soara,  Romania}}

\maketitle

\begin{abstract}
A Lagrangian theory giving rise to a  version of the Dirac-K\" ahler
equations on curved backgrounds is considered. The principal pieces
are the general fields which have values in the algebra of the Dirac
matrices and satisfy a Dirac-type equation.  Their components are
scalar, pseudo-scalar, vector, axial-vector fields and fields
strength which satisfy an irreducible systems of first-order
Dirac-K\" ahler equations having remarkable gauge and duality
properties similar to those of the flat case. The vector and
axial-vector fields are the physical potentials giving rise to the
field strength while the scalar fields play an auxiliary role and
can be eliminated by fixing a suitable gauge. The chiral components
of the field strength are either self-dual or anti self-dual with
respect to the Hodge duality.

Pacs: 04.62.+v
\end{abstract}


\newpage

\section{Introduction}

The theory of the relativistic quantum fields focuses especially on
the spinor and vector fields describing the fundamental particles
(quarks, leptons and gauge bosons) and on the scalar field which is
a universal tool or substitute in theories on flat or curved
background. The free fields satisfy the well-known Klein-Gordon,
Proca and Dirac equations which look like having different origins
as long as these are of different orders. This conjecture encouraged
many authors to seek for first-order equations, known as Bhabha
equations \cite{Bha},  even for the fields with integer spins.
Successful attempts are the Duffin-Kemmer-Petiau  theory
\cite{DKP}-\cite{DKP2}, \cite{KN} and some recent generalizations
\cite{R,LPT,KTBR} based on  special algebras that are completely
different to that of the Dirac matrices \cite{DM,SW1}. However,  the
natural generalization the Dirac theory is the Dirac-K\"ahler
approach \cite{Kah} which gives rise to first-order equations for
systems of scalar vector and tensor fields \cite{Krug} we call here
{\em dual systems}.

In the present paper we would like to study the Lagrangian theory of
the dual systems on curved manifolds by using a simple algebraic
method \cite{Krug,Spehler} restricting ourselves to consider only
free fields minimally coupled to the gravity of the curved
spacetime. We exploits the fact that the Dirac matrices, which form
a basis of a 16-dimensional algebra over ${\Bbb C}$, behave as
scalars, vectors or skew-symmetric tensors under the transformations
of the {\em gauge} group of any local-Minkowskian spacetime
\cite{SG}. Therefore, we start with general fields defined on curved
manifolds with values in the algebra of the Dirac matrices
\cite{Krug}. These fields have to satisfy the usual free Dirac
equation (in a matrix version) corresponding to the minimal coupling
to gravity. In this manner we obtain the matrix form of the
Dirac-K\" ahler equations \cite{Dfor} of the dual systems. The
methods of the Dirac theory enable us to study the specific features
of this approach in local frames or in natural ones recovering the
known properties of the flat case \cite{Krug}.

The resulted system of the first-order Dirac-K\" ahler equations is
irreducible and remarkably coherent. The vector and axial-vector
fields play the role of potentials giving rise to a skew-symmetric
field strength. Other equations couple either the scalar and the
vector fields to the field strength or the pseudo-scalar and the
axial-vector to the dual field of this field strength (in the sense
of the Hodge duality \cite{MTW}). It is interesting that, in an
arbitrary gauge, the vector and axial-vector fields satisfy the
second-order Proca equation but without to accomplish the Lorentz
condition which is mandatory in the Proca theory. This is because of
the special position of the scalar  fields which take over the
contribution of the divergences of the vector fields being involved
in their gauge transformations \cite{Spehler}. Under such
circumstances, it is obvious that only the vector and axial-vector
fields have a specific physical meaning. In addition, these
equations have interesting property of chirality and duality
resulted from the Dirac-K\" ahler theory.

This paper is organized as follows. In section 2 we briefly present
the Dirac formalism on curved spacetimes pointing out the role of
the 16-dimensional algebra of the point-dependent Dirac matrices in
local frames.  In the next section we define the general fields with
values in this algebra and we build the Lagrangian theory generating
the Dirac-K\" ahler equation for the dual system of fields. Herein
we derive the second order equation and we study the duality
properties using the methods of the Dirac theory. The section 4 is
devoted to the covariant formalism in which we present the final
form of the Lagrangian density and the resulted Dirac-K\" ahler
field equations whose properties are investigated. In section 5 we
discuss the chirality and duality properties pointing out that the
chiral components of the field strength are either self-dual or anti
self-dual. Finally we present our conclusion while in Appendix we
give the algebra we use here.

\section{The Dirac formalism}

The theory of the fields with half-integer spins on curved
local-Mikowskian manifolds, $(M,g)$, can be formulated in any local
chart $\{x\}$, of coordinates $x^{\mu}$ ($\mu,\nu,...=0,1,2,3$), but
only in local (non-holonomic) frames, $(e)$, defined by the tetrad
fields $e_{\hat\mu}$ and $\hat e^{\hat\mu}$. These fields are
labeled by {\em local} indices, $\hat\mu,\hat\nu,...=0,1,2,3$, which
are raised or lowered by the Minkowski metric $\eta={\rm
diag}(1,-1,-1,-1)$ while for the natural ones we have to use the
metric tensor $g_{\mu \nu}=\eta_{\hat\alpha\hat\beta}\hat
e^{\hat\alpha}_{\mu}\hat e^{\hat\beta}_{\nu}$. Therefore, the
general geometric context of the theories with spin is given by
$(M,g)$ and $(e)$ \cite{SG,BD}.

The Dirac theory deals with the 16-dimensional algebra ${\cal
A}/{\Bbb C}$ of the complex - valued $4\times 4$ matrices where we
consider the basis
$\{I,\gamma^5,\gamma^{\hat\mu},\gamma^{\hat\mu}\gamma^5,S^{\hat\mu\hat\nu}\}$.
The usual Dirac matrices satisfy $\{
\gamma^{\hat\mu},\gamma^{\hat\nu}\}=2\tilde\eta^{\hat\mu\hat\nu}I$
where $I$ is the identity matrix. The matrices
$S^{\hat\mu\hat\nu}=\frac{i}{4}\,[\gamma^{\hat\mu},\gamma^{\hat\nu}]$
are the basis generators of the spinor representation
$\rho_s=(\frac{1}{2},0)\oplus (0,\frac{1}{2})$ of the subgroup ${\rm
Spin}(\eta)=SL(2,{\Bbb C})\subset {\rm Spin}(\tilde\eta)$ which is
the {\em gauge} group of $(M,g)$. This plays an important role since
the {\em covariant} fields on $(M,g)$ transform under isometries
according to the representations of the isometry group {\em induced}
by the finite-dimensional representations $\rho$ of the gauge group
\cite{ES,EPL}. For this reason, the algebra ${\cal A}$ becomes the
principal algebraic ingredient in constructing the tetrad-gauge
covariant theory of the fields with spin on  $(M,g)$.

The Lagrangian theory of the free Dirac field $\psi$ of mass $m$,
minimally coupled to the gravity of $(M,g)$, is based on the action
${\cal S}_D=\int d^4 x\sqrt{g}\,  {\cal L}_D$ where we denote
$g=|\det(g_{\mu\nu})|$. For given $(M,g)$ and $(e)$, the Lagrangian
density reads \cite{BD,CD1},
\begin{equation}
{\cal L}_D=
\frac{i}{2}\,[\overline{\psi}\gamma^{\hat\alpha}\nabla_{\hat\alpha}\psi-
(\overline{\nabla_{\hat\alpha}\psi})\gamma^{\hat\alpha}\psi] -
m\overline{\psi}\psi\,,\quad \overline{\psi}=\psi^+\gamma^0\,.
\end{equation}
This depends on the covariant derivatives
$\nabla_{\hat\alpha}=e_{\hat\alpha}^{\mu}\nabla_{\mu}$ whose action
on the spinor field,
$\nabla_{\mu}\psi=(\partial_{\mu}+\Gamma^{spin}_{\mu})\psi$, is
given by the spin connections
\begin{equation}
\Gamma_{\mu}^{spin}=\hat
e_{\mu}^{\hat\alpha}\,\Gamma_{\hat\alpha}^{spin}=\frac{i}{2}\,
e^{\beta}_{\hat\nu} (\hat
e^{\hat\sigma}_{\alpha}\Gamma^{\alpha}_{\beta\mu}- \hat
e^{\hat\sigma}_{\beta,\mu} )
S^{\hat\nu\,\cdot}_{\cdot\,\hat\sigma}\,,
\end{equation}
which satisfy $\overline{\Gamma}_{\mu}^{spin}=-\Gamma_{\mu}^{spin}
$. The Dirac equation
$(i\gamma^{\hat\alpha}\nabla_{\hat\alpha}-m)\psi=0$, resulted from
the action ${\cal S}_D$, is tetrad-gauge invariant since  the
covariant derivatives assure the covariance of the whole theory
under the tetrad-gauge transformations,
\begin{eqnarray}
\hat e^{\hat\alpha}_{\mu}(x)&\to& \hat e'^{\hat\alpha}_{\mu}(x)=
\Lambda^{\hat\alpha\,\cdot}_{\cdot\,\hat\beta}[A(x)]
\,\hat e^{\hat\beta}_{\mu}(x)\nonumber\\
e_{\hat\alpha}^{\mu}(x)&\to&  {e'}_{\hat\alpha}^{\mu}(x)=
\Lambda_{\hat\alpha\,\cdot}^{\cdot\,\hat\beta}[A(x)]
\,e_{\hat\beta}^{\mu}(x)\label{gauge}\\
\psi(x)&\to&~\psi'(x)=T[A(x)]\,\psi(x)\nonumber
\end{eqnarray}
determined by  the local transformations $A(x)\in SL(2,\Bbb C)$. The
transformation matrices of the spinor representation $\rho_s$ are
denoted by $T(A)$ while the notation $\Lambda (A)$ stands for the
Lorentz transformations which correspond to the $SL(2,{\Bbb C})$
ones through the canonical homomorphism \cite{W}. The matrices
$\Lambda (A)$ transform the quantities carrying local indices - for
example, the Dirac matrices transform as
$\overline{T}(A)\gamma^{\hat\alpha}{T(A)}=
\Lambda^{\hat\alpha\,\cdot}_{\cdot\,\hat\beta}(A)\gamma^{\hat\beta}$
(since $\overline{T}=T^{-1}$).

At this stage it is convenient to introduce in each local frame
$(e)$ the point-dependent matrices
\begin{equation}\label{gese}
\gamma^{\mu}(x)=e^{\mu}_{\hat\alpha}(x)\gamma^{\hat\alpha}\,,\quad
S^{\mu\nu}(x)=e^{\mu}_{\hat\alpha}(x)e^{\nu}_{\hat\beta}(x)
S^{\hat\alpha\hat\beta}\,,
\end{equation}
which have the advantage to be  {\em covariantly constant}
(commuting with the covariant derivatives) \cite{SG}. Without a
mathematical close-up view, we note that, in each point $x\in(M,g)$,
the set
$\{I,\gamma^5,\gamma^{\mu}(x),\gamma^{\mu}(x)\gamma^5,S^{\mu\nu}(x)\}$
defined by Eq. (\ref{gese}) represents a basis of the algebra ${\cal
A}[e(x)]$ which obeys the usual algebraic rules but with $g(x)$
replacing $\eta$, as presented in Appendix. Based on such properties
one derives the commutation relations of the covariant derivatives
\cite{SG},
\begin{equation}\label{Com1}
[\nabla_{\mu},\,\nabla_{\nu}]\psi= \textstyle\frac{1}{4}
R_{\alpha\beta\mu\nu}\gamma^\alpha\gamma^{\beta}\psi \,,
\end{equation}
and the identity $R_{\alpha\beta\mu\nu}\gamma^{\beta}
\gamma^{\mu}\gamma^{\nu}=-2R_{\alpha\nu}\gamma^{\nu}$ where
$R_{\alpha\beta \mu\nu}$ is  the curvature tensor and
$R_{\alpha\beta}=R_{\alpha \mu \beta\nu}g^{\mu\nu}$ . Hereby one
deduces the second-order equation,
\begin{equation}\label{mass}
(i\gamma^{\mu}\nabla_{\mu}+m)(i\gamma^{\nu}\nabla_{\nu}-m)\psi=
-\left(\nabla^2-\textstyle{\frac{1}{4}}R+m^2\right)\psi=0
\end{equation}
(with $\nabla^2=g^{\mu\nu}\nabla_{\mu}\nabla_{\nu}$ and
$R=R_{\mu\nu}g^{\mu\nu}$) playing the same role as the Klein-Gordon
mass condition in special relativity.

\section{Dirac-K\" ahler equations in matrix form}

We use the operator formalism of the Dirac theory exploiting the
properties of the algebra ${\cal A}$ for constructing the Lagrangian
theory of the vector fields on $(M,g)$. We consider the {\em dual
system} of complex-valued fields, $(f,h,A,B,F) : (M,g)\to {\Bbb C}$,
formed by the vector $A$ and the axial-vector $B$ which play the
role of potentials generating the field strength
$F_{\mu\nu}=-F_{\nu\mu}$. The scalar $f$ and the pseudo-scalar  $h$
are auxiliary fields involved in the gauge transformations of the
fields $A$ and respectively $B$. For this reason we say that the
pair $(f,A)$ represents the {\em vector} sector while the fields
$(h,B)$ form the {\em axial} sector. This system is called dual
since  the vector sector will be related to the field $F$ while the
axial sector will couple its {\em dual} field, ${^*F}$, which has
the components
$^*F_{\alpha\beta}=\frac{1}{2}\,\tilde\varepsilon_{\alpha\beta\mu\nu}F^{\mu\nu}$.

Given a dual system, we define the {\em general field} $W : (M,g)
\to {\cal A}$ in each local frame $(e)$ as
\begin{equation}\label{W}
W=ifI+h\gamma^5+A_{\hat\mu}\gamma^{\mu}-i
B_{\nu}\gamma^{\nu}\gamma^5 +F_{\mu\nu}S^{\mu\nu}\,.
\end{equation}
As in the flat case \cite{Krug}, the principal piece of our proposal
is the action ${\cal S}_W=\int d^4 x\sqrt{g}\, {\cal L}_W$, of the
field $W$ {\em minimally} coupled to the gravity of $(M,g)$, assumed
to have  the following Lagrangian density,
\begin{equation}\label{Lag}
{\cal L}_W=-\frac{1}{4}{\rm Tr}\left\{
\frac{i}{2}\,\left[\overline{W}\gamma^{\alpha}\nabla_{\alpha}W-
(\overline{\nabla_{\alpha}W})\gamma^{\alpha}W\right] -
m\overline{W}W\right\}\,,
\end{equation}
where $\overline{W}=\gamma^0 W^+\gamma^0$ is the Dirac adjoint of
$W$. The spinor covariant derivatives are defined now by the rule
\begin{equation}\label{nabW}
\nabla_{\mu}W=\partial_{\mu}W+\left[ \Gamma_{\mu}^{spin},W
\right]\,,
\end{equation}
corresponding to the gauge transformations,
\begin{equation}\label{gW}
W(x)\to W'(x)=T[A(x)]W(x)\overline{T}[A(x)]\,,
\end{equation}
of the fields $W$ which have two spinor indices (as the matrix
$\psi\overline{\psi}$).  Furthermore, considering the matrix
elements of the fields $W$ and $\overline{W}$ as the canonical
variables of the Lagrangian density (\ref{Lag}), after a few
manipulation, we find the Dirac-K\" ahler equation in matrix form
\cite{Dfor},
\begin{equation}\label{eqW}
i\gamma^{\alpha}\nabla_{\alpha}W-m W=0\,,
\end{equation}
where the covariant derivatives act as in (\ref{nabW}). This is in
fact a system of 16 first-order differential equations which are
linear independent, determining thus all the 16 components of the
dual system $(f,h,A,B,F)$ we consider here. In this approach we have
the advantage to find easily the commutator of the covariant
derivatives,
\begin{equation}
[\nabla_{\mu},\,\nabla_{\nu}]W = \textstyle\frac{1}{4}
R_{\alpha\beta\mu\nu}[\gamma^\alpha\gamma^{\beta}, W] \,,
\end{equation}
by using Eqs. (\ref{Com1}) and (\ref{nabW}). Then we can deduce the
second-order equations by multiplying Eq. (\ref{eqW}) with its
Klein-Gordon divisor. We obtain the general second-order equation
\begin{eqnarray}
&&-(i\gamma^{\mu}\nabla_{\mu}+m)(i\gamma^{\nu}\nabla_{\nu}-m)W
\nonumber\\
&&\hspace*{18mm} = (\nabla^2+m^2)W-\textstyle{\frac{i}{2}}
R_{\alpha\beta\mu\nu}S^{\alpha\beta}[S^{\mu\nu},W]=0 \label{mass}
\end{eqnarray}
which gives rise to the mass conditions of all the fields
$(f,h,A,B,F)$ we present in the next section.

Hence we derived the Dirac-K\" ahler equations in matrix form  which
can be studied using the methods of the Dirac theory that  help us
to investigate the properties of the field equations and the effects
of the internal or space-time symmetries. Because of the special
structure of ${\cal L}_W$ the symmetry transformations may have the
general form  $W\to W'=UW\overline{V}$ where $\overline{U}=U^{-1}$
and $\overline{V}=V^{-1}$.  We remind the reader that the internal
symmetries preserve the Lagrangian density but the space-time
isometries can not do this leaving merely the action invariant.
Obviously, these symmetries depend on the geometry of $(M,g)$ and
the couplings of the concrete physical model. Here we consider the
simplest example of  the $U(1)$ transformations $W\to We^{i\theta}$
depending on the point-independent parameter $\theta \in [0,2\pi)$.
As in the case of the Dirac field, the conserved quantity
corresponding to this symmetry is the conserved current
\begin{equation}\label{J}
J^{\mu}=\frac{1}{4}\, {\rm
Tr}\left(\overline{W}\gamma^{\mu}W\right)\,,\quad
J^{\mu}_{\,\,;\mu}=0\,.
\end{equation}

Another opportunity of the operator approach is the analysis of the
chiral projections of the general fields \cite{Spehler},
\begin{equation}\label{chW}
W=(P_L+P_R)W(P_L+P_R)=W_{RR}+W_{LL}+W_{RL}+W_{LR}\,,
\end{equation}
obtained with the help of the projection matrices
$P_{L,R}=\frac{1}{2}\,(I\mp \gamma^5)$. Bearing in mind that the
operator $i\gamma^{\mu}\nabla_{\mu}$ has only $LR$ and $RL$ blocks
we can split the field equation in four chiral projections,
\begin{eqnarray}
&&(i\gamma^{\mu}\nabla_{\mu})_{LR}W_{RR}=mW_{LR}\,,\quad
(i\gamma^{\mu}\nabla_{\mu})_{RL}W_{LR}=mW_{RR}\,,\label{ch1}\\
&&(i\gamma^{\mu}\nabla_{\mu})_{LR}W_{RL}=mW_{LL}\,,\quad
(i\gamma^{\mu}\nabla_{\mu})_{RL}W_{LL}=mW_{RL}\,,\label{ch2}
\end{eqnarray}
which can be helpful in some applications. These projections have
different behaviors under duality transformations since the {\em
dual} field of $W$ is defined as $^*W=iW\gamma^5$ and satisfies
\begin{equation}\label{WW}
^*W(f,h,A,B,F)=W(h,-f,B,-A,{^*F})\,.
\end{equation}
Moreover, taking into account that ${^*(^*F)}=-F$, we deduce a
similar property for the general fields, ${^*(^*W)}=-W$. This
conjecture enables us to introduce the {\em duality rotations}
\cite{MTW},
\begin{equation}\label{chrot}
W\to W'= W e^{i\theta_{ch}\gamma^5}\,,\quad \theta_{ch}\in
[0,2\pi)\,,
\end{equation}
whose effect is to multiply Eqs. (\ref{ch1}) with $e^{i\theta_{ch}}$
and Eqs. (\ref{ch2}) with $e^{-i\theta_{ch}}$ but without to change
the solutions. Nevertheless,  these transformations can not be
associated to an internal symmetry since they change the structure
of ${\cal L}_W$. For example, taking $\theta_{ch}=\frac{\pi}{2}$ we
have $W'={^*W}$ and ${\cal L}_{^*W}=-{\cal L}_{W}$ understanding
that the fields $W$ and $^*W$ satisfy the same equation (\ref{eqW})
but derived from different actions.

\section{Lagrangian formalism}

The operator approach helped us to find the field equations and to
do a rapid inspection of their properties. Now we must rewrite this
theory exclusively in the covariant Lagrangian formalism as long as
there are only fields carrying natural indices. This can be done (in
usual notations, $\nabla_{\mu}=\, _{;\mu}$ and $\partial_{\mu}=\,
_{,\mu}$) since the covariant derivatives of the general fields,
\begin{equation}\label{DW}
\nabla_{\sigma}W=if_{\,,\sigma}I+h_{\,,\sigma}\gamma^5+A_{\mu\,;\sigma}\gamma^{\mu}-i
B_{\nu\,;\sigma}\gamma^{\nu}\gamma^5
+F_{\mu\nu\,;\sigma}S^{\mu\nu}\,,
\end{equation}
may be expressed in terms of the covariant derivatives the fields
$(f,h,A,B,F)$. Assuming  that the components of these fields and
their complex conjugated fields,
$(\bar{f},\bar{h},\bar{A},\bar{B},\bar{F})$ represent the new
canonical variables,  we replace first Eqs. (\ref{W}) and (\ref{DW})
in Eq. (\ref{Lag}). Then, according to the properties listed in
Appendix, we find the definitive form of the Lagrangian density,
\begin{eqnarray}
{\cal L}_W&=&~\,
m\left(\bar{f}f-\bar{h}h+\bar{A}^{\mu}A_{\mu}-\bar{B}^{\mu}B_{\mu}+
\textstyle{\frac{1}{2}}\,\bar{F}^{\mu\nu}
F_{\mu\nu}\right)\nonumber\\
&&+\textstyle{\frac{1}{2}}\,\left(\bar{A}^{\mu}f_{\,,\mu}+\bar{f}_{\,,\mu}A^{\mu}
-\bar{A}^{\mu}_{\,;\mu}f
-\bar{f}A^{\mu}_{\,;\mu}\right)\nonumber\\
&&-\textstyle{\frac{1}{2}}\left(\bar{B}^{\mu}h_{\,,\mu}+\bar{h}_{\,,\mu}B^{\mu}
-\bar{B}^{\mu}_{\,;\mu}h
-\bar{h}B^{\mu}_{\,;\mu}\right)\nonumber\\
&&+\textstyle{\frac{1}{2}}\left(\bar{A}_{\mu\,;\nu}F^{\mu\nu}+\bar{F}^{\mu\nu}A_{\mu\,;\nu}
-\bar{A}_{\mu}F^{\mu\nu}_{~~;\nu} -\bar{F}^{\mu\nu}_{~~;\nu}A_{\mu}
\right)\nonumber\\
&&-\textstyle{\frac{1}{2}}\left(\bar{B}_{\mu\,;\nu}{^*F^{\mu\nu}}
+{^*\bar{F}}^{\mu\nu}B_{\mu\,;\nu}
-\bar{B}_{\mu}\,{^*F}^{\mu\nu}_{~~;\nu}
-{^*\bar{F}}^{\mu\nu}_{~~;\nu}B_{\mu}\right)\label{Lagu}\,.
\end{eqnarray}
First of all, we remark that this  is {\em irreducible} since the
fields of the vector and axial sectors (which appear here with
opposite signs) can not be separated among themselves because of the
field strength which couples both the vector and axial-vector
fields. Nevertheless, a whole sector can be eliminated by dropping
out simultaneously both the scalar and vector fields of this sector.

The Lagrangian density (\ref{Lagu}) gives rise to the following {\em
covariant} Dirac-K\" ahler system of first-order equations
\begin{eqnarray}
&&A^{\mu}_{~;\mu}-m\,f=0\,,\qquad
~~F^{\,\mu}_{\,\,\cdot\,\sigma\,;\mu}+f_{\,,\sigma}+m\, A_{\sigma}=0\,,\label{eq1}\\
&&B^{\mu}_{~;\mu}-m\,h=0\,,\qquad
{^*F}^{\,\mu}_{\,\,\cdot\,\sigma\,;\mu}+h_{\,,\sigma}+m\,
B_{\sigma}=0\,,\label{eq2}\\
&&~~~-\tilde\varepsilon_{\,\mu\nu}^{\,\,\,\cdot\,\cdot\,\sigma\tau}B_{\sigma\,;\tau}
+A_{\mu\,;\nu}-A_{\nu\,;\mu}+m\, F_{\mu\nu}=0\,,\label{eq3}\\
&&~~~~~~\tilde\varepsilon_{\,\mu\nu}^{\,\,\,\cdot\,\cdot\,\sigma\tau}A_{\sigma\,;\tau}
+B_{\mu\,;\nu}-B_{\nu\,;\mu}+m\,{^*F}_{\mu\nu}=0\,,\label{eq4}
\end{eqnarray}
where Eqs. (\ref{eq1}), (\ref{eq2}) and (\ref{eq3}) represent the
Euler-Lagrange equations  deduced from (\ref{Lagu}) while Eq.
(\ref{eq4}) is derived from (\ref{eq3}). We specify that Eqs.
(\ref{eq1})-(\ref{eq3}) are linear combinations of those given by
Eq. (\ref{eqW}) which means that both these systems are equivalent
\cite{Dfor}.

The coefficients of Eqs. (\ref{eq1})-(\ref{eq4}) are exclusively
real numbers thanks to our special parametrization (\ref{W}).
Consequently, there is a particular case in which all our fields can
have real-valued components, $f,h,A,B,F\in {\Bbb R}$. We observe
that in this case we must change the Lagrangian density taking only
a half of the contribution given by Eq. (\ref{Lagu}). However, in
general, we have to consider complex-valued fields producing the
non-vanishing conserved current,
\begin{eqnarray}
J_{\mu}&=&i\left(\bar{A}_{\mu}f-\bar{f}A_{\mu}+\bar{A}^{\nu}F_{\mu\nu}
-\bar{F}_{\mu\nu}A^{\nu}\right.\nonumber\\
&&\left.-\bar{B}_{\mu}h+\bar{h}B_{\mu}-\bar{B}^{\nu}\,{^*F}_{\mu\nu}
+\,{^*\bar{F}}_{\mu\nu}B^{\nu}\right)\,,
\end{eqnarray}
resulted from Eq. (\ref{J}).

The second-order equations deduced from Eq. (\ref{mass}) or derived
directly by using  Eqs. (\ref{eq1})-(\ref{eq4}) have the form
\begin{equation}\label{mass1a}
\left(\nabla^2+m^2\right)\left|
\begin{array}{c}
f\\
h
\end{array}\right|=0 \,,\quad \left(\nabla^2+m^2\right)\left|
\begin{array}{c}
A_{\mu}\\
B_{\mu}
\end{array}\right|
-R_{\mu\,\cdot}^{\,\,\,\cdot\,\nu}\left|
\begin{array}{c}
A_{\nu}\\
B_{\nu}
\end{array}\right|=0
\end{equation}
The first equations indicate that $f$ and $h$ are Klein-Gordon
fields of mass $m$. Moreover, Eqs. (\ref{mass1a}b)  coincide to that
of a Proca field $X$, of the same mass, obeying the mandatory
Lorentz condition $X^{\mu}_{\,\,;\mu}=0$. What is new here is that
our fields ${A}$ and $B$ satisfy the Proca equation without
accomplishing the Lorentz condition. This is possible because of the
scalar fields ${f}$ and $h$ which take over the contributions of the
divergences  $A^{\mu}_{~;\mu}$ and $B^{\mu}_{~;\mu}$ and couple
between themselves the scalar and vector sectors in Eqs.
(\ref{eq1}b) and (\ref{eq2}b).

On the other hand, the scalar fields can be modified according to
our current needs since Eqs. (\ref{eq1})-(\ref{eq4}) remain
invariant under the {\em gauge} transformations \cite{Spehler},
\begin{eqnarray}
f\to  f'= f-\alpha \,,&\quad& A_{\mu}\to
A_{\mu}'=A_{\mu}+\frac{1}{m}\,\partial_{\mu}\alpha\,,\label{gau1}\\
h\to  h'= h-\alpha \,,&\quad& B_{\mu}\to
B_{\mu}'=B_{\mu}+\frac{1}{m}\,\partial_{\mu}\beta\,,\label{gau2}
\end{eqnarray}
where the scalar fields $\alpha,\,\beta : (M,g)\to {\Bbb C}$ satisfy
the Klein-Gordon equation, $(\nabla^2+m^2)(\alpha,\beta)=0$. It is
worth pointing out that this gauge affects the scalar fields which
can be eliminated by taking $\alpha=f$ and $\beta=h$. For example,
the gauge-fixing  $\alpha={f}$ drops out the field $f$ and imposes
the Lorentz condition ${ A}^{\mu}_{~;\mu}=0$ so that the vector
field $A$ becomes a genuine complex-valued Proca field but
contributing to the field strength given by Eq. (\ref{eq3}).
However, whether in this gauge we vanish the axial sector taking, in
addition, $h=0$ and $B=0$ then we remain only with the Proca field
$A$ and its traditional field strength which satisfy the well-known
equations $A_{\mu\,;\nu}-A_{\nu\,;\mu}+m\, F_{\mu\nu}=0$,
$F^{\,\mu}_{\,\,\cdot\,\sigma\,;\mu}+mA_{\sigma}=0$ and
${^*F}^{\,\mu}_{\,\,\cdot\,\sigma\,;\mu}=0$.

The general conclusion is that only two fields may have a specific
physical significance. These are the vector field $A$ and the
axial-vector field $B$ which play the role of potentials generating
the field strength $F$. The scalar fields can not have an
independent meaning since their form depend on the gauge-fixing.

\section{Self-duality}

The dual system  $(f,h,A,B,F)$  includes pairs of independent scalar
and vector fields but with different behaviors under the parity
transformation. This situation discourage us to consider arbitrary
linear combinations of these fields apart from some special cases of
physical (or mathematical) interest as that of the chiral
projections defined by Eq. (\ref{chW}). We observe that these can be
expressed in a simpler form as,
\begin{eqnarray}
&W_{LL}=i{_{(+)}\!{ f}}P_L+\textstyle{\frac{1}{2}}\,{_{(+)}\!{
F}}_{\mu\nu} S^{\mu\nu}\,,\quad &W_{RL}={_{(+)}\!{
A}}_{\mu}\gamma^{\mu}P_L\,,
\label{Wch1}\\
&W_{RR}=i{_{(-)}\!{ f}}P_R+\textstyle{\frac{1}{2}}\,{_{(-)}\!{
F}}_{\mu\nu} S^{\mu\nu}\,, \quad &W_{LR}={_{(-)}\!{
A}}_{\mu}\gamma^{\mu}P_R\,,\label{Wch2}
\end{eqnarray}
marking out the new fields
\begin{equation}\label{fAFch}
_{(\pm)}\!{  f}=f\pm ih\,,\quad _{(\pm)}\!{A}=A\pm iB\,,\quad
_{(\pm)}\!{ F}=F\pm i\,{^*F}\,,
\end{equation}
which mix the vector and axial sectors. Consequently,  the parity
transformation has to change the $(\pm)$ components into the $(\mp)$
ones just as in the case of the chiral components of the Dirac
field. Therefore, we can say that Eqs. (\ref{fAFch}) define the {\em chiral}
components of our dual system.

Changing the parametrization again, we consider the fields
(\ref{fAFch}) and their complex conjugated fields, ${_{(\pm)}\!\bar{
f}},\, {_{(\pm)}\!\bar{ A}}$, and ${_{(\pm)}\!\bar{ F}}$, as the new
canonical variables of the Lagrangian density (\ref{Lagu}). Then we
find that the system of equations (\ref{eq1})-(\ref{eq4}) splits in
two independent chiral subsystems, denoted by $(+)$ and $(-)$, that
read
\begin{eqnarray}
&&_{(\pm)}\!{  A}^{\mu}_{~;\mu}-m\,_{(\pm)}\!{ f}=0\,,\quad
~~_{(\pm)}\!{  F}^{\,\mu}_{\,\,\cdot\,\sigma\,;\mu}+ {_{(\pm)}\!{
f}}_{\,,\sigma}+m\,
{_{(\pm)}\!{  A}}_{\sigma}=0\,,\label{eq11}\\
&&~~~\pm
i\,\tilde\varepsilon_{\,\mu\nu}^{\,\,\,\,\cdot\,\cdot\,\sigma\tau}
{_{(\pm)}\!{  A}}_{\sigma\,;\tau} +{_{(\pm)}\!{  A}}_{\mu\,;\nu}
-{_{(\pm)}\!{  A}}_{\nu\,;\mu}+m\, {_{(\pm)}\!{
F}}_{\mu\nu}=0\,.\label{eq12}
\end{eqnarray}
This splitting is rather formal, at the level of the field equations
only, since the Lagrangian density still mixes the components of the
chiral subsystems $(\pm)$. For example, the mass term of the scalar fields
becomes now
\begin{equation}\label{ffhh}
\bar{f}f-\bar{h}h \to \textstyle{\frac{1}{2}}\left( {_{(+)}\! \bar{
f}}{_{(-)}\! { f}}+ {_{(-)}\! \bar{ f}}{_{(+)}\! {  f}}\right)\,.
\end{equation}
We note that Eqs. (\ref{eq11}) and (\ref{eq12}) can be obtained
directly by replacing Eqs. (\ref{Wch1}) and (\ref{Wch1}) in Eqs.
(\ref{ch1}) and (\ref{ch2}). Moreover, the second-order equations
are similar to Eqs. (\ref{mass1a}) while the gauge transformations
(\ref{gau1}) and (\ref{gau2}) can be put in the compact form,
\begin{equation}
{_{(\pm)}\!{  f}}\to {_{(\pm)}\!{  f}}'={_{(\pm)}\!{
f}}-{_{(\pm)}\!{\alpha}} \,,\quad {_{(\pm)}\!{  A}_{\mu}}\to
{_{(\pm)}\!{  A}}_{\mu}'={_{(\pm)}\!{
A}}_{\mu}+\frac{1}{m}\;\partial_{\mu}{_{(\pm)}\!{\alpha}}\,,
\end{equation}
denoting ${_{(\pm)}\!{\alpha}}=\alpha+i\beta$.

We obtained thus two simple chiral subsystems  we expect to have
remarkable duality properties. Indeed, bearing in mind that the
fields strength and their complex conjugated fields satisfy
\begin{equation}
{^*_{(\pm)}\!{F}}=\mp \, i\, {_{(\pm)}\!{  F}}\,, \quad
{^*_{(\pm)}\!\bar{F}}=\pm i\, {_{(\pm)}\!\bar{  F}}\,,
\end{equation}
we draw the conclusion that the fields ${_{(-)}\!{  F}}$ and
${_{(+)}\!\bar{  F}}$ are {\em self-dual} while ${_{(+)}\!{  F}}$
and ${_{(-)}\!\bar{  F}}$ are {\em anti self-dual}. In the
particular case of real-valued fields, $f,h,A,B,F\in {\Bbb R}$, the
supplemental rule ${_{(\pm)}\!\bar{F}}={_{(\mp)}\!{F}}$ holds for
all the fields of the subsystems $(\pm)$ and we are left only with
one self-dual field, ${_{(-)}\!{  F}}={_{(+)}\!\bar{F}}$, and one
anti-self dual field, ${_{(+)}\!{F}}={_{(-)}\!\bar{F}}$.  We must
specify that the structure of Eqs. (\ref{eq12}) is crucial in
generating the duality properties.

Under such circumstances, the duality rotations (\ref{chrot})
transform the fields $(\pm)$ as,
\begin{eqnarray}
&&{_{(\pm)}\!{  f}}\to {_{(\pm)}\!{  f}}e^{\mp
i\theta_{ch}}\,,~~\quad
_{(\pm)}\!{  A}\to _{(\pm)}\!{ A} e^{\mp i\theta_{ch}}\,,\nonumber\\
&&{_{(\pm)}\!{  F}}\to {_{(\pm)}\!{ F}}\cos\theta_{ch}\pm
{^*_{(\pm)}\!{ F}}\sin\theta_{ch}={_{(\pm)}\!{  F}}e^{\mp
i\theta_{ch}}\,,
\end{eqnarray}
with different phase factors which do not modify the solutions of
the systems (\ref{eq1})-(\ref{eq2}) but affect the form of ${\cal
L}_W$. We may verify this fact simply, observing that the mass term
$(\ref{ffhh})$ transforms as
\begin{equation}\label{ffhh1}
{_{(+)}\! \bar{f}}{_{(-)}\! { f}}+ {_{(-)}\! \bar{ f}}{_{(+)}\! {
f}}\to e^{2i\theta_{ch}}{_{(+)}\! \bar{f}}{_{(-)}\! { f}}+
e^{-2i\theta_{ch}}{_{(-)}\! \bar{ f}}{_{(+)}\! {  f}} \,.
\end{equation}
Concluding we can say that the chiral subsystems are self-dual since
their field strength are either self-dual or anti self-dual and, in
addition, the duality rotations do not mix components of different
signs $(\pm)$.

\section{Concluding remarks}

The Dirac theory offered us the appropriate framework for analyzing
the first-order  Dirac-K\" ahler equations governing the dual
systems minimally coupled to the gravity of a curved background.
Each system has only two fields of physical relevance, the vector
and axial-vector fields. These have the same mass but this does not
represent a redundance as long as these fields play different
physical roles. More specific, in a concrete physical model the
vector field couples a vector current while the axial-vector field
must be coupled to an axial current. We note that these couplings
may increase the coherence of the model when the vector and
axial-vector fields are related to each other as components of the
same dual system \cite{Spehler}.

However, before to study the physical behavior of the dual systems,
there are many technical problems to be solved. The first one is
that of the parametrization for which we do not have  many options
if we keep the gauge-covariant Dirac-type equations. Nevertheless, a
new parametrization must be introduced in the massless case
\cite{mass0} since in the present one the field strength is cut out
from its potentials when the mass vanishes . Further investigation
may focus on the most interesting part of this theory concerning the
structure and properties of the stress-energy tensor and the
conserved quantities corresponding to the internal or space-time
symmetries. A surprise could be to find that the dual systems on
curved backgrounds deal with models of dark mater or energy.

\subsection*{Acknowledgements}
I would like to thank Matej Pavsic for drawing me attention on a
crucial part of the bibliography.

\subsection*{Appendix: The algebras ${\cal A}[e(x)]$}

For given $(M,g)$ and $(e)$, the matrices
$\{I,\gamma^5,\gamma^{\mu},\gamma^{\mu}\gamma^5,S^{\mu\nu}\}$ form a
basis of the mapping ${\cal A}(e): (M,g)\to {\cal A}$. In each point
$x\in(M,g)$ the algebra ${\cal A}[e(x)]$ is isomorphic to ${\cal A}$
according to Eqs. (\ref{gese}). This means that any pair of
algebras, ${\cal A}[e(x)]$ and ${\cal A}[e'(x')]$, are isomorphic to
each other. Fortunately, the algebraic rules of these algebras do
not depend on $(e)$ but only on the metric tensor $g$. The
commutation rules are
\begin{eqnarray}
&[\gamma^{\mu},\gamma^{\nu}]=-4iS^{\mu\nu}\,,\quad&[S^{\mu\nu},\gamma^{5}]\,=\,0\,, \\
&[\gamma^{\mu},\gamma^{\nu}\gamma^5]=2g^{\mu\nu}\gamma^5\,,\quad
&[S^{\mu\nu},\gamma^{\sigma}]=i(g^{\nu\sigma}\gamma^{\mu}
-g^{\mu\sigma}\gamma^{\nu})\,, \\
&[\gamma^{\mu}\gamma^5,\gamma^{\nu}\gamma^5]=4iS^{\mu\nu}\,, \quad
&{[}S^{\mu\nu},\gamma^{\sigma}\gamma^5{]}
=i(g^{\nu\sigma}\gamma^{\mu}\gamma^5
-g^{\mu\sigma}\gamma^{\nu}\gamma^5)\,,
\end{eqnarray}
\begin{equation}
{[}S^{\mu\nu},S^{\sigma\tau}{]}=i(
g^{\mu\tau}S^{\nu\sigma}-g^{\mu\sigma}S^{\nu\tau}+g^{\nu\sigma}S^{\mu\tau}
-g^{\nu\tau}S^{\mu\sigma})\,,
\end{equation}
while the anti-commutation ones read
\begin{eqnarray}
&\{\gamma^{\mu},\gamma^{\nu}\}=2g^{\mu\nu}I\,,\quad
&\{S^{\mu\nu},\gamma^{5}\}=-i\,\tilde\varepsilon\,^{\mu\nu\,\cdot\,\cdot}
_{\,\cdot\,\cdot\,\sigma\tau}\,S^{\sigma\tau}\,,\\
&\{\gamma^{\mu},\gamma^{\nu}\gamma^5\}=-2\,\tilde\varepsilon\,^{\mu\nu\,\cdot\,\cdot}
_{\,\cdot\,\cdot\,\sigma\tau}\,S^{\sigma\tau}\,,\quad
&\{S^{\mu\nu},\gamma^{\sigma}\}=\tilde\varepsilon\,^{\mu\nu\sigma\,\cdot}
_{\,\cdot\,\cdot\,\cdot\,\,\tau}
\,\gamma^{\tau}\gamma^5\,,\\
&\{\gamma^{\mu}\gamma^5,\gamma^{\nu}\gamma^5\}=-2g^{\mu\nu}I\,,\quad
&\{S^{\mu\nu},\gamma^{\sigma}\gamma^5\}
=\tilde\varepsilon\,^{\mu\nu\sigma\,\cdot}_{\,\cdot\,\cdot\,\cdot\,\,\tau}
\,\gamma^{\tau}\,,
\end{eqnarray}
\begin{equation}
\{S^{\mu\nu},S^{\sigma\tau}\}=\frac{1}{2}\,
(g^{\mu\sigma}\eta^{\nu\tau}-g^{\nu\sigma}g^{\mu\tau})I-\frac{i}{2}\,
\tilde\varepsilon\,^{\mu\nu\sigma\tau}\gamma^5\,,
\end{equation}
where we denote
$\tilde\varepsilon\,^{\mu\nu\sigma\tau}=e^{\mu}_{\hat\alpha}\,e^{\nu}_{\hat\beta}
\,e^{\sigma}_{\hat\gamma}\,e^{\tau}_{\hat\delta}\,
\varepsilon^{\hat\alpha\hat\beta\hat\gamma\hat\delta}$ \cite{MTW}
adopting the convention $\varepsilon_{0123}=-\varepsilon^{0123}=1$
for the usual Levi-Civita symbol
$\varepsilon_{\hat\alpha\hat\beta\hat\gamma\hat\delta}$ carrying
local indices. In addition we use the identities \cite{MTW}
\begin{equation}
\tilde\varepsilon\,_{\mu\nu\sigma\tau}\,\tilde\varepsilon\,^{\alpha\beta\sigma\tau}=
-2\left(\delta^{\alpha}_{\mu}\delta^{\beta}_{\nu}-\delta^{\alpha}_{\nu}\delta^{\beta}_{\mu}
\right)\,, \quad
\tilde\varepsilon\,_{\mu\nu\sigma\tau}\,\tilde\varepsilon\,^{\alpha\nu\sigma\tau}
=-6\,\delta^{\alpha}_{\mu}\,.
\end{equation}

\end{document}